\documentclass[twocolumn]{revtex4}

\usepackage{graphicx}
\usepackage{epsfig}
\usepackage{amsmath, amssymb}
\usepackage{verbatim}
\usepackage{bm}

\renewcommand{\b}[1]{\boldsymbol{#1}}
\newcommand{\gammabar}{\ensuremath\gamma\kern-0.53em-}

\begin{document}

\title{Transitions Between Chiral Spin Liquids and $Z_2$ Spin Liquids}
\author{Maissam Barkeshli}
\affiliation{Department of Physics, Stanford University, Stanford, CA 94305 }

\begin{abstract}
The Kalmeyer-Laughlin chiral spin liquids (CSL) and the $Z_2$ spin liquids are two of the simplest 
topologically ordered states. Here I develop a theory of a direct quantum phase
transition between them. Each CSL is characterized by an integer $n$ and is topologically equivalent to the 
$1/2n$ Laughlin fractional quantum Hall (FQH) state. Depending on the parity of $n$, the transition
from the CSL is either to a ``twisted'' version of the $Z_2$ spin liquid, the ``doubled semion'' model, or
the conventional $Z_2$ spin liquid (``toric code''). 
In the presence of $SU(2)$ spin symmetry, the triplet gap remains open through the transition and only 
singlet operators acquire algebraic correlations. An essential observation is that the CSL/Laughlin FQH states can be understood in terms of bosonic 
\it integer \rm quantum Hall  (BIQH) states of Schwinger bosons or vortices,
respectively. I propose several novel many-body wave functions that can interpolate through the transition. 
\end{abstract}

\maketitle

Two of the simplest quantum spin liquids in two spatial dimensions that exhibit topological order \cite{wen04,nayak2008} 
are the Kalmeyer-Laughlin (KL) chiral spin liquid (CSL) \cite{kalmeyer1987,wen1989}
and the $Z_2$ quantum spin liquid \cite{lee2006}.
These states played significant roles in the development of the theory of topological order and in attempts to formulate a theory
of high $T_c$ superconductivity \cite{lee2006}. It has been established that these states can be stabilized with local 
Hamiltonians \cite{kitaev2003} 
and there is  increasingly strong evidence from numerical simulations that physically realistic Hamiltonians for frustrated magnets \cite{bauer2013, jiang2012a,jiang2012b,yan2011,depenbrock2012} or 
bosons in partially filled Chern bands \cite{sorenson2005} 
can also stabilize them \footnote{Hamiltonians
with longer ranged interactions have also been found to stabilize the KL-CSLs \cite{schroeter2007}.
Other CSLs have also been found as ground states of exactly solvable models \cite{kitaev2006,yao2007}, 
although these are in a different universality class as compared with the KL-CSL studied in this paper.}. 
However it has not been clear whether a direct, continuous quantum phase transition is possible between the 
Laughlin and $Z_2$ states. Aside from being of intrinsic interest, understanding such questions may be helpful 
for the challenge of identifying the nature of topological spin liquids, such as those found in frustrated Heisenberg 
models \cite{jiang2012a,jiang2012b,yan2011,depenbrock2012},  or in 
interpreting experimental observations of gapless spin liquids \cite{lee2008}.

The CSL was argued to be realized in frustrated spin-$1/2$ systems, which can mapped to a problem of bosons 
in a magnetic field via a Holstein-Primakoff transformation \cite{kalmeyer1987}. 
The bosons in turn form a $1/2n$ Laughlin fractional quantum Hall (FQH) state.
Later it was shown \cite{wen1989} that the KL-CSL, for $n = 1$, can also be understood
in terms of a projective construction where fermionic spinons form a fermionic integer quantum Hall state. 
In contrast, the $Z_2$ spin liquid is described at low energies by $Z_2$ gauge theory, due to pair condensation
of spinons. There are two topologically distinct classes of $Z_2$ gauge theory \cite{dijkgraaf1990} 
and thus two corresponding classes of $Z_2$ spin liquids; as will be reviewed below, these describe the ``toric code'' 
and ``doubled-semion'' models \cite{lee2006,kitaev2003}.

\begin{figure}
\includegraphics[width=3.4in]{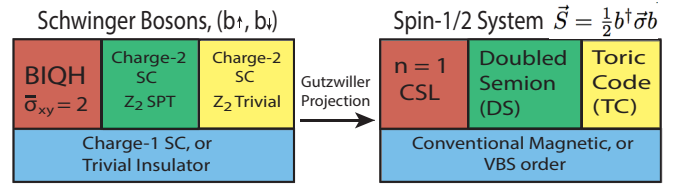}
\caption{\label{globalPD}Possible global phase diagram for an $SU(2)$ invariant spin-1/2 system. 
The triplet gap remains open at the CSL-DS transition, but closes at the DS-TC transition. 
If $SU(2)$ is broken, both the DS-TC transition and the CSL-VBS transition can each split into
two transitions. See (\ref{H}) for a possible Hamiltonian. 
}
\end{figure}

In this paper, I develop a theory of a direct continuous transition between the CSL and the $Z_2$ states.
For odd $n$, the transition is to the doubled-semion model, 
and for even $n$, it is to the toric code model. In the Appendix I also provide a theory of the transition 
between the doubled-semion and toric code models, which is shown to be described by QED$_3$. 
A key insight is that, in contrast to previous constructions, the CSL can be understood as a state 
where \it Schwinger bosons \rm form a bosonic \it integer \rm quantum
Hall (BIQH) state with Hall conductance $\bar{\sigma}_{xy} = 2n$. 
The BIQH state possesses no fractionalized excitations, and is
topologically non-trivial only in the presence 
of a conserved $U(1)$ charge \cite{senthil2013,lu2012}. 
It forms a simple example of a bosonic symmetry-protected topological
(SPT) phase \cite{chen2013,senthil2013,lu2012,vishwanath2013,levin2012,swingle2012,cheng2013}. 
The bosonic Laughlin FQH state can be understood as a state where the vortices of a bosonic superfluid collectively form a BIQH state. 
Using these observations, I show that the transitions between the Laughlin and $Z_2$ 
states can be understood in terms of transitions between BIQH states and charge-$2$ superconductors (SC) (see Fig. \ref{globalPD}).
The latter transitions can be shown to be in the XY universality class. Depending on the parity of $n$, the 
charge-$2$ SC can itself be mapped to a topologically non-trivial Ising paramagnet, of the type discussed in \cite{levin2012}. 

\it{BIQH states and transitions} --\rm A bosonic system can form an IQH state with no fractionalized 
excitations, as long as the electrical Hall conductance is quantized as 
$\sigma_{xy}/\sigma_0 = 2n \equiv \bar{\sigma}_{xy}$, where $n$ is an integer and $\sigma_0 = Q^2/h$ is the appropriate conductance 
quantum for bosons of charge $Q$ (In the following we set $\sigma_0 = 1/2\pi$.)\cite{senthil2013}. 
This can be described by the following effective $U(1) \times U(1)$ Chern-Simons (CS) gauge theory:
\begin{align}
\label{BIQHLag1}
\mathcal{L} = \frac{1}{4\pi} K_{IJ} a^I \partial a^J + \frac{1}{2\pi} A_{e} \partial (a^1 + a^2),
\end{align}
with $K = \left(\begin{matrix} 0 & 1 \\ 1 & 2(1-n) \end{matrix} \right)$, where $A_{e}$ is the external electromagnetic gauge field,
and where $a^I \partial a^J \equiv \epsilon^{\mu \nu \lambda} a^I_\mu \partial_\nu a^J_\lambda$. 
This theory has counterpropagating edge states and therefore no net thermal Hall effect, and no topologically non-trivial
excitations (since $|\text{Det } K|  =1$). However the charge is carried by one of the chiral edge modes, leading to the 
quantized $\sigma_{xy}$: $U(1)$ charge conservation prohibits backscattering and thus protects the 
counterpropagating edge modes \cite{senthil2013}. 

Let us focus on the case $n  =1$.  This state can be understood in terms of composite fermions
and reverse flux attachment \cite{jainCF}, applied to bosons at filling fraction $\nu  =2$. 
To each boson we attach $-2\pi$ flux of a statistical gauge field, leading to composite fermions at 
filling factor $\nu_{cf} = -2$, which then form the $\nu_{cf} =-2$ IQH state. This composite fermion construction can be alternatively understood 
by writing the hard-core bosons in terms of two fermionic ``partons'' : $b = f_1 f_2$, which introduces a compact $U(1)$ gauge field 
$a$ associated with the gauge redundancy $f_{1} \rightarrow e^{i\theta} f_1$, $f_2 \rightarrow e^{-i\theta} f_2$. We 
consider a mean-field state of the partons where $f_1$, $f_2$ each form a quantized Hall insulator with Chern number
$C_1 = -1$ and $C_2 = 2$, respectively. In this language, $f_2$ plays the role of the ``composite fermion,'' while $a$ plays the role of the 
emergent statistical gauge field. When the gauge fluctuations are properly included (see Appendix \ref{biqhProjConApp}), 
this mean-field ansatz yields a BIQH state with no fractionalization, described by the field theory (\ref{BIQHLag1}).

The BIQH wave function suggested by this construction is given by
\begin{align}
\label{bqhwfn}
\Psi_{bqh}(\{\b{r}_i\}) = \Phi_{-1}(\{\b{r}_i\}) \Phi_{2}(\{\b{r}_i\}) ,
\end{align}
where $\Phi_{C}(\{ \b{r}_i\} )$ is a Slater determinant wave function for fermions forming a quantized Hall insulator with
total Chern number $C$. A simple example of (\ref{bqhwfn}) for a continuum system at filling fraction 
$\nu =2$ is given by $\Psi_{bqh}(\{z_i\}) = \prod_{i < j} (z_i - z_j)\Phi_{-2}(\{z_i^*\})$, where now
$\Phi_{-2}$ is a free fermion wave function for two filled Landau levels at $\nu = -2$. 

\begin{table}
\begin{tabular}{cccl}
\hline
$(C_1, C_2)$ & Resulting state & Hall conductance, $\sigma_{xy}/\sigma_0$\\
\hline
$(-1,2)$ & BIQH & $2$ \\
$(-k,k)$ & Charge-$k$ SC ($Z_k$ SPT)  & --\\
$(-1,0)$ & Trivial Mott insulator & $0$\\
$(-1,-1)$ & $1/2$ -Laughlin & $1/2$ \\
\hline\\
\end{tabular}
\caption{
\label{statesTable}
Chern number assignments for the mean-field insulating states of partons 
$f_1$ and $f_2$, and resulting bosonic states, where $b = f_1 f_2$. The charge-$k$
SC breaks $U(1)$ charge conservation and does not have a well-defined quantized DC 
Hall conductivity; if its Goldstone mode is gapped by, eg a long-ranged interaction, then it describes a $Z_k$ SPT. 
}
\end{table}

The advantage of the above parton construction is that other, nearby collective states of the bosons
can also be understood by considering mean-field states associated with other values of the Chern numbers (see Table \ref{statesTable}). 
The continuous transitions between these various states can therefore be described by the
Chern number changing transition for $f_2$ or $f_1$, coupled to the emergent
$U(1)$ gauge field $a$ \cite{barkeshli2012sf,grover2013,lu2013}. 

For the current discussion, consider the case $(C_1, C_2) = (-k, k)$. We will show that the resulting state of the bosons is a charge-$k$
superconductor and, moreover, is closely related to a \it topologically non-trivial \rm $Z_k$ SPT state (the case $k =1$ has
been described previously \cite{barkeshli2012sf,grover2013,lu2013,laughlin1988}). 
To see this, let $f_1$ and $f_2$ have charges $0$ and $1$, respectively, under an external gauge field $A_{e}$, such that $b$ has 
charge $1$. Integrating out $f_1$, $f_2$ yields the following Lagrangian density
\begin{align}
\label{sfLag}
\mathcal{L} 
= \frac{k}{2\pi} A_{e} \partial a + \frac{k}{4\pi} A_{e} \partial A_{e} 
- \frac{1}{g^2}(\epsilon_{\mu\nu\lambda}\partial_\nu a_\lambda)^2 + \cdots,
\end{align}
where we have included a higher order term for $a$; 
in general, all higher order terms compatible with gauge invariance of $a$ and $A_{e}$ will be allowed. 
Remarkably, (\ref{sfLag}) implies that instantons of $a$ carry charge $k$, and therefore are 
prohibited from proliferating at low energies due to charge conservation. Consequently, $a$ is effectively a massless noncompact
gauge field. (\ref{sfLag}) shows that the fluctuations of $a$ are coupled to $A_e$ and 
therefore correspond to physical density/current fluctuations. Thus this theory must describe a superfluid state, and 
$a$ is the dual of the Goldstone mode. Since we consider the bosons to carry charge, we will use the terms
superconductor and superfluid interchangeably.

In this theory, the vortices of the superfluid correspond to the gapped fermionic excitations of the
Chern bands, as these are minimally coupled to the dual Goldstone mode $a$. 
To see that this describes a charge-$k$ superconductor, suppose $A_e$ is dynamical, so
that the dual Goldstone mode $a$ is gapped by the Anderson-Higgs mechanism. 
The low energy theory is
$\mathcal{L} = \frac{k}{2\pi} a \partial A_{e} + a \cdot (j_1 - j_2) + \cdots$,
where we have included the currents $j_1$ and $j_2$ associated with the partons $f_1$ and $f_2$, and
the $\cdots$ indicate higher derivative terms. 
The equation of motion for $a$ is now a constraint:
$\epsilon_{\mu \nu \lambda} \partial_\nu A_{e;\lambda} = \frac{2\pi}{k} (j_{2;\mu} - j_{1;\mu})$.
This directly shows that the gapped excitations associated with $f_1$ and $f_2$ each carry $2\pi/k$ 
units of flux, as expected for the vortices of a charge-$k$ superconductor. 
Another way to see this is to introduce a field $\xi_\mu = \frac{1}{2\pi} \epsilon^{\mu \nu \lambda} \partial_\nu a_\lambda$, 
and a Lagrange multiplier field $\varphi$ to enforce $\partial_\mu \xi_\mu = 0$. 
Integrating out $\xi_\mu$ then gives the Lagrangian of a charge-$k$ superconductor:
 $\mathcal{L} \propto (\partial \varphi - k A_e)^2$.

In the case where $A_e$ is dynamical so that $a$ is gapped by the Anderson-Higgs mechanism, 
$b$ therefore forms a charge-$k$ gapped superconductor with $Z_k$ symmetry. Remarkably, this 
state is a topologically non-trivial $Z_k$ SPT phase. From the group cohomology classification \cite{chen2013}, there are $k$ 
topologically distinct bosonic SPT states with $Z_k$ symmetry.
In a topologically trivial $Z_k$ state, gauging the $Z_k$ symmetry will yield ordinary
$Z_k$ gauge theory, the topologically properties of which can be described by a $U(1) \times U(1)$ CS theory with
$K$-matrix $K = \left(\begin{matrix} 0 & k \\ k & 0 \end{matrix} \right)$. This is sometimes also referred to as 
the $Z_k$ toric code model: there are $k^2$ topologically distinct quasiparticles $(a,b)$, for $a,b = 0,...,k-1$,
with statistics $\theta_{(a,b)} = 2\pi ab/k$. However, if we consider the external
response field $A_e$ to be a dynamical gauge field in (\ref{sfLag}), then we obtain a $U(1) \times U(1)$ CS theory with 
$K = \left(\begin{matrix} 0 & k \\ k & k \end{matrix} \right)$. The topological properties of such a theory
are distinct from those of $Z_k$ toric code. 
\footnote{It is crucial for the discussion that we assigned integer charges to $f_1$ and $f_2$.
This is required in order to have the conventional periodicities for the large gauge transformations of the $U(1) \times U(1)$ CS
gauge fields, $\oint_C a^I \cdot dl\sim \oint_C a^I \cdot dl + 2\pi$ for non-contractible loops $C$.}. 
The case $k = 2$ describes the so-called doubled semion model, which consists of four
particles, $1, s, \bar{s}, s\times \bar{s}$, with statistics $\theta_s = \theta_{\bar{s}} = \pi/2$, 
$\theta_{s \times\bar{s}}= \pi$, and with $s^2 = \bar{s}^2 = 1$. Therefore the bosons
$b$ must be forming a non-trivial $Z_k$ SPT state \cite{levin2012}.

The transition between the BIQH state with $\bar{\sigma}_{xy} = 2$ and the charge-$2$ SC can now be understood
as a transition where $f_1$ changes Chern number from $C_1 =-1 \rightarrow -2$, while $f_2$ remains inert. 
This suggests a wave function that can continuously interpolate between the two phases:
\begin{align}
\label{bqhsfwfn}
\Psi_{bqh - Z_2}(m;\{ \b{r}_i \})= \Phi_{f1}(m; \{ \b{r}_i \}) \Phi_{2}( \{ \b{r}_i \}),
\end{align}
where $\Phi_{f1}(m; \{ \b{r}_i \})$ is a Slater determinant wave function for the $f_1$ state, which tunes from
a quantized Hall insulator with $C_1 = -1$ to $-2$ as the tuning parameter $m$ is tuned through $0$. 

The field theory for this transition is given by 
\begin{align}
\label{xyFerm}
\mathcal{L} = &\frac{1}{2} \frac{1}{4\pi} a \partial a + \bar{\psi} \gamma^\mu (\partial_\mu - i a_\mu) \psi + m \bar{\psi} \psi
\nonumber \\
&+ \frac{2}{2\pi} A_{e} \partial a + \frac{2}{4\pi} A_{e} \partial A_{e} ,
\end{align}
where $\psi$ is a two-component Dirac fermion associated with the Chern number changing transition of $f_1$,
$\bar{\psi} \equiv \psi^\dagger \gamma^0$, and $\gamma^\mu$ are the Pauli matrices. 
When $m < 0$, integrating out the Dirac fermion $\psi$ cancels the CS term for $a$, 
giving (\ref{sfLag}). When $m > 0$, integrating out $\psi$ yields the BIQH state. The action above can be recognized as
the fermionization of the 3D XY transition \cite{chen1993,barkeshli2012sf} (see Appendix \ref{xyAppendix}), and therefore is 
dual to the following Chern-Simons-Higgs theory:
\begin{align}
\label{xyBIQH}
\mathcal{L} = |(\partial - i k A_{e}) \Phi|^2 + m |\Phi|^2 + \lambda |\Phi|^2 - \frac{2n}{4\pi} A_{e} \partial A_{e},
\end{align}
with $(k,n) = (2,1)$, $\lambda > 0$, and $\Phi$ a complex scalar field. For general $(k,n)$, it is clear that (\ref{xyBIQH}) describes a transition between a 
bosonic state with $\bar{\sigma}_{xy} =2n$ and a charge-$k$ superconductor. The critical theory (\ref{xyBIQH}) is
expected on general grounds: By definition, the bulk of an SPT phase is trivial, and so any $XY$ ordering transition 
in the bulk of a BIQH state should be conventional. The above derivation using the parton construction provides both 
a highly non-trivial check and projected many-body wave functions. When $A_e$ in (\ref{xyBIQH}) is interpreted 
as an internal dynamical gauge field, the $\Phi$-condensed phase describes the doubled semion theory (see Appendix \ref{CSHapp}), 
agreeing with the parton construction analysis. 

The $n=1$ BIQH state can also be understood in terms of two-component bosons, 
$b_\alpha$, where the spin index $\alpha = \uparrow , \downarrow$. In the parton construction we set 
$b_\alpha = f_1 f_\alpha$, and consider $f_\alpha$ to form a spin singlet $\nu  =-2$ IQH state (\it ie \rm $f_\alpha$
form Chern insulators with Chern number $C_\alpha = 1$), while again $f_1$ forms a quantized Hall 
insulator with $C_1 = -1$. This construction was used in \cite{senthil2013} to describe the 
BIQH state, and also earlier in \cite{barkeshli2010twist}
to study topological phase transitions. This suggests the following wave function for a continuum
system: $\Psi(\{z_i^\uparrow,z_i^\downarrow\}) = \prod_{i < j} |z_i^\uparrow - z_j^\uparrow|^2 
|z_i^\downarrow - z_j^\downarrow|^2 \prod_{i,j} (z_i^{\uparrow *} - z_j^{\downarrow *}) e^{-\sum_{i,\alpha} |z_i^\alpha|^2/4 l_B^2}$
where $l_B$ is the magnetic length. The generalization to 
a lattice system is straightforward. Note that this is a \it spin singlet \rm wave function. 
In the two-component case, the transition to the charge-$2$ SC is again understood 
as a transition where $C_1$ changes from $-1$ to $-2$. Since the spinful fermions $f_\alpha$ are not modified, 
$SU(2)$ spin symmetry can be preserved and the state can remain a spin singlet throughout the transition. 

\it{CSL/Laughlin FQH states in terms of BIQH states} \rm -- 
Consider an $SU(2)$ invariant spin-1/2 system, and write the spin-1/2 operator in terms of Schwinger bosons as
$\vec{S} = \frac12 b^\dagger \vec{\sigma} b$,
where $b^\dagger = (b^\dagger_\uparrow, b^\dagger_\downarrow)$
is a two-component complex scalar boson. This description introduces a $U(1)$ gauge field $a$
associated with the gauge transformation $b \rightarrow e^{i\theta} b$, which keeps physical 
operators invariant and implements the constraint $b^\dagger b = 1$. 
The description of the conventional magnetically ordered or valence bond solid (VBS) states in terms of Schwinger 
bosons (Fig. \ref{globalPD}) follows from previous work\cite{lee2006}. 
Here we consider a mean-field state where $b$ forms the spin singlet BIQH state with $\sigma_{xy} = 2$. 
Using the effective field theory (\ref{BIQHLag1}) 
for the BIQH state, it follows that the effective theory for the CSL is given by (\ref{BIQHLag1}), but
with $A_{e}$ replaced by the dynamical gauge field $a$,
and where the current of $b_\alpha$ is given by 
$j_{\alpha;\mu} = \frac{1}{2\pi} \epsilon^{\mu \nu \lambda} \partial_\nu a^\alpha_\lambda$. 
Integrating out $a^\alpha$ then gives the known theory for the KL-CSL \cite{wen1989,wen04}:
$\mathcal{L} = -\frac{2}{4\pi} a \partial a + a \cdot (j_\uparrow + j_\downarrow)$,
where the currents $j_\alpha$ for the gapped bosons $b_\alpha$ are now included in the low energy theory 
as ``test charges.'' 

If instead the mean-field band dispersion of $b_\alpha$ has $n$ minima at momenta
$\b{Q}_\beta$ for $\beta = 1, ..., n$, then we can approximate $b_\alpha(\b{r}) \approx \sum_{\beta  = 1}^n e^{i \b{Q}_\beta \cdot \b{r}} b_{\alpha\beta}(\b{r})$,
leading to $n$ flavors of two-component bosons $b_{\alpha \beta}$. 
A mean-field state where each of $n$ flavors forms a spin singlet $\bar{\sigma}_{xy} = 2$ BIQH state then leads, at long wavelengths, to
the effective theory $\mathcal{L} = -\frac{2n}{4\pi} a \partial a + a \cdot \sum_{\alpha \beta} j_{\alpha \beta}$, where
$j_{\alpha \beta}$ is the current of $b_{\alpha \beta}$. This theory has $2n$ topologically distinct quasiparticle excitations with fractional statistics
$\theta_{k} = \pi k^2/n \text{ mod } 2\pi$, and therefore is topologically equivalent to the original KL-CSL. 
In contrast, the fermionic spinon construction of \cite{wen1989}, where the fractional statistics are given by 
$\theta_{k} = \pi(k^2/n + k) \text{ mod } 2\pi$, has a \it different \rm topological order when $n > 1$. 
Thus for all $n$ the construction we have presented is in the same universality class as the state 
originally proposed by KL, but not to the state proposed in \cite{wen1989}. The reason is that \cite{wen1989}
starts with a construction of fermionic spinons, $\vec{S} = \frac12 c^\dagger \vec{\sigma} c$, and considers a mean-field state
where the fermionic spinons $c$ form a $\nu  =2n$ \it fermionic \rm IQH state, leading to the extra shifts of $\pi$ in $\theta_k$. 
Within this mean-field theory it is not possible to understand the continuous transition to the $Z_2$ spin liquid,
which necessitates the construction of this paper (see Appendix \ref{slaveFermApp}). 

The above construction suggests that the wave function for the CSL can be obtained from the 
Gutzwiller projection of the BIQH state of $b$ onto the physical Hilbert space of one particle per site:
\begin{align}
\label{wfnPg}
|\Psi_{phys} \rangle = \mathcal{P}_{G} |\Phi_{BIQH}\rangle,
\end{align}
where $|\Phi_{BIQH}\rangle$ is the mean-field state describing the BIQH state of $b$, and 
$\mathcal{P}_{G}$ is the Gutzwiller projection. (\ref{wfnPg}) is in general different 
from the wave functions proposed originally \cite{kalmeyer1987,wen1989}.  

Now consider bosons with a conserved $U(1)$ charge in the FQH regime, instead of $SU(2)$ invariant spin-1/2 
systems. The vortices of a Bose superfluid can be described in terms of a bosonic field $\phi_v$ coupled to a non-compact
gauge field $A$, where the particle current is $j_\mu = \frac{1}{2\pi} \epsilon_{\mu\nu\lambda} \partial_\nu A_\lambda$. 
This implies that the vortices see the original particles as a magnetic field.
When bosons are at filling fraction $\nu = 1/2n$, the vortices are at an effective filling fraction $\nu_v = 2n$, and therefore are poised to form a 
BIQH state. The effective field theory of the vortices is given by
$\mathcal{L} = \frac{1}{2\pi} A_{e} \partial A + A \cdot j_v + \mathcal{L}_v(\phi_v)$, where 
$\mathcal{L}_v(\phi_v)$ is the effective theory of the vortices and $j_v$ is the vortex current. 
When the vortices form the BIQH state, described by the effective theory (\ref{BIQHLag1}),
we obtain a $U(1)^3$ CS gauge theory described by a $K$-matrix 
$K = \left(\begin{matrix} 0 & 1 & 1 \\ 1 & 2(n-1) & 1 \\ 1 & 1 & 0 \end{matrix} \right)$.
This satisfies $|\text{Det } K|  = 2n$, and the fractional statistics of the quasiparticles coincide with those of the
$1/2n$ Laughlin state. This can be seen more directly by integrating out $a^1$ and $a^2$ to obtain
$\mathcal{L} = -\frac{2n}{4\pi} A \partial A + \frac{1}{2\pi} A_e \partial A$, which is the more conventional
theory of the $1/2n$ Laughlin FQH state \cite{wen04}. 

It follows that when the vortices form either the BIQH state, a trivial Mott insulator, or a superfluid, then the original
bosons form either the Laughlin FQH state, the superfluid, or the trivial Mott insulator, respectively. Therefore the
former transitions, studied in \cite{grover2013,lu2013}, when interpreted in terms of vortices, describe the latter transitions, studied in
\cite{barkeshli2012sf}.

Let us turn to wave functions. A simple trial wave function for a single vortex in a continuum bosonic superfluid 
takes the form \cite{onsager1949}:
$\Psi_v(\eta;\{\b{r}_i \}) = \prod_i (z_i - \eta) g(|z_i - \eta|/\xi )\Psi_0(\{\b{r}_i \})$,
where $z_i = \b{r}_{i;x} + i \b{r}_{i;y}$ is the complex coordinate of the $i$th boson, $\eta$ is the complex coordinate
of the vortex, and $g(r) \propto 1/r$ is a real function that allows for the relaxation of the particle density profile of the vortex
in the interacting superfluid. The ground state wave function $\Psi_0$ of the interacting superfluid is usually considered to take
the Jastrow form $\Psi_0(\{\b{r}_i \}) = e^{ \sum_{i < j} u(r_i - r_j)}$, for a suitable real function $u(r)$.
This suggests that the many-body boson wave function for a state where the vortices have formed
a BIQH state may be described by the following ansatz:
\begin{align}
\label{vortexFQHwfn}
\Psi_{fqh}( \{ \b{r}_i\}) = \int \prod_l d^2\eta_l \Psi_v(\{\eta_j\}, \{\b{r}_i \}) \Phi_{bqh}(\{\eta_j\}) 
\end{align}
where $\Phi_{bqh}(\{\eta_i\})$ is the BIQH wave function for particles with complex coordinates $\{\eta_i\}$,
and $\Psi_v(\{\eta_j\}, \{\b{r}_i \})$ is the natural generalization of 
$\Psi_v(\eta;\{\b{r}_i \})$ to a state with many well-separated vortices. 
(\ref{vortexFQHwfn}) and its connection to the effective field theory studied here is reminiscent of
the standard FQH hierarchy construction \cite{wen04}.

\it{Continuous transitions between CSL/Laughlin FQH states and $Z_2$ fractionalized states} \rm --
Let us begin with the case of the $SU(2)$ invariant spin liquid. When the Schwinger
bosons form the BIQH state with $\bar{\sigma}_{xy} = 2$, the spin-1/2 system forms the $n = 1$ KL-CSL
state. When the Schwinger bosons form the charge-$2$ SC described above,
then the arguments given previously imply that the resulting spin liquid is topologically
equivalent to the ``doubled semion'' model. The critical theory is given by (\ref{xyBIQH}), with 
$A_{e}$ instead replaced by the emergent dynamical gauge field $a$. 
A wave function that interpolates through the transition can be written by using (\ref{wfnPg}), and with
$|\Phi_{BIQH}\rangle$ replaced by the spinful wave function of the BIQH to charge-2 SC transition described
above. 
The generalization to $n > 1$ is given in Appendix \ref{generalN}. Note that in the above construction,
the spins remain gapped through the transition: only spin singlet operators appear in the critical theory.
More details about the physical operators at the critical point are discussed in Appendix \ref{physicalOps}. 

A possible microscopic Hamiltonian that may tune through this transition is the following
$SU(2)$ symmetric spin-$1/2$ model on the Kagome lattice:
\begin{align}
\label{H}
H = \lambda H_{csl} + H_{Z_2},
\end{align}
where $H_{csl} =J_c \sum_{i,j,k\in \bigtriangleup, \bigtriangledown} \vec{S}_i \cdot (\vec{S}_j \times \vec{S}_k)$ with the
sum over the triangles of the Kagome lattice, with $i,j,k$ ordered clockwise around the vertices of the triangles,
and $H_{Z_2} = J_1 \sum_{\langle i j \rangle } \vec{S}_1 \cdot \vec{S}_2 + J_2 \sum_{\langle \langle ij \rangle \rangle} \vec{S}_1 \cdot \vec{S}_2$
\cite{bauer2013b}. 
$H_{CSL}$ has recently been shown to lead to a gapped $n = 1$ CSL,\cite{bauer2013} and $H_{Z_2}$ has recently been shown
to lead to a topological spin liquid either in the toric code or doubled semion universality classes
\cite{yan2011,jiang2012a,depenbrock2012} when $J_2/J_1$ is close to zero. 
The considerations of this paper suggest that a natural transition out of the $n =1$ 
CSL is to the doubled semion model; the simplest possible phase diagram of (\ref{H}) is therefore that
when $\lambda \ll 1$, $H$ realizes the doubled semion model, and when $\lambda \gg 1$ it realizes
the CSL, and there is a direct continuous transition between them when $\lambda \sim 1$. 
More complicated phase diagrams are also possible; in Appendix \ref{dstcSec} we
will provide a theory of a transition between the doubled semion and toric code theories, which 
may also appear in the global phase diagram (Fig. \ref{globalPD}). 

In the case of bosons with a conserved $U(1)$ charge, the transition between the Laughlin FQH state
and the $Z_2$ states can be understood as a transition where the vortices
undergo a BIQH to charge-$2$ SC transition, which is again given by (\ref{xyBIQH}), 
with $A_{e}$ instead replaced by the noncompact gauge field $A$ to which the vortices are coupled. Now,
a wave function that interpolates through the transition is given by (\ref{vortexFQHwfn}), with 
$\Phi_{bqh}(\{\eta_i\})$ replaced by the BIQH - charge-$2$ SC transition, (\ref{bqhsfwfn}). 

\it{Acknowledgments}\rm -- I thank X.-L. Qi, X.-G. Wen, T. Senthil, N. Read, Y. Zhang, Z. Wang, J. McGreevy
T. Grover, B. Swingle and especially S. Kivelson, Bela Bauer and Meng Cheng for discussions 
and/or comments on the manuscript. 
I also thank J. McGreevy, C. Laumann, and N. Yao for collaborations on related work, and the Simons Center for 
Geometry and Physics program on Topological States of Matter for hospitality
while part of this work was completed. This work was supported by the Simons Foundation. 

\appendix

\section{Continuous transition between doubled semion and toric code models}
\label{dstcSec}

Here we note that it is possible to understand a continuous transition between two kinds of 
$SU(2)$ invariant $Z_2$ spin liquids, corresponding to the doubled semion and toric code models, respectively. 
Specifically, the doubled semion model can continuously transition to the toric code model in the presence of $SU(2)$ spin symmetry. 
The Lagrangian at long wavelengths is described by $QED_3$ with two flavors of Dirac fermions ($N_f = 2$). 
Remarkably, in this case the gaps of both triplet and singlet excitations approach zero at the quantum phase transition. 
In the absence of $SU(2)$ spin symmetry, there can be generically an intervening topologically trivial gapped insulator
unless other symmetries are present to stabilize the direct transition. 

To see this, first consider the BIQH phase diagram studied in \cite{grover2013,lu2013}. This can be understood
using the following field theory:
\begin{align}
\label{biqhtrans}
\mathcal{L} = &\sum_{k=\uparrow, \downarrow} \bar{\psi}_k [\gamma^\mu (\partial_\mu - i \alpha_\mu) + M_k] \psi_k - \frac{1}{g} (\epsilon_{\mu\nu\lambda}\partial_\nu \alpha_\lambda)^2
\nonumber \\
&-\frac{1}{4\pi} A_e \partial A_e - \frac{1}{2\pi} A_e \partial \alpha,
\end{align}
where here $\alpha$ is an emergent $U(1)$ gauge field, $\psi_k$ for $k  =\uparrow, \downarrow$ are each 
two-component Dirac fermions, and $k$ is a physical $SU(2)$ spin index. 
$\gamma^\mu$ are the Pauli matrices, and $\bar{\psi}_k \equiv \psi_k^\dagger \gamma^0$. 
$SU(2)$ symmetry implies that $M_\uparrow = M_\downarrow$. 
When $M_1 = M_2 > 0$, the theory is in the BIQH state; when $M_1 = M_2 < 0$, the theory is
in a topologically trivial insulating state with $\bar{\sigma}_{xy} = 0$. When $M_1$ and $M_2$ have opposite
signs, which breaks the $SU(2)$ spin symmetry, then the theory can be shown to be in a superfluid state. 
\begin{figure}
\centerline{
\includegraphics[width=3.2in]{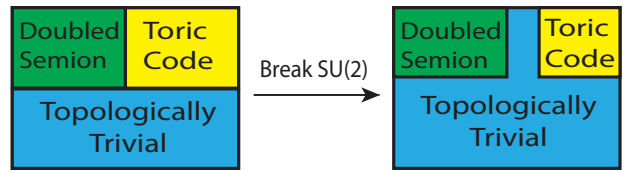}
}
\caption{\label{dstcFig}
Possible schematic phase diagram for $Z_2$ spin liquids. A direct transition between the doubled semion and 
toric code models can occur in the presence of $SU(2)$ spin symmetry. The topologically trivial phase may itself 
spontaneously break $SU(2)$ and correspond to a conventional magnetically ordered state.
In the absence of $SU(2)$ spin symmetry, the direct transition splits into two transitions, 
with the topologically trivial insulating phase intervening. 
}
\end{figure}
(\ref{biqhtrans}) can be understood as arising from the following parton construction
\begin{align}
b_k = f_0 f_k,
\end{align}
where $k = \uparrow,\downarrow$ and $b_k$ have charge $1$ under the external gauge field $A_e$. 
Next, we assume a mean-field ansatz where $f_0$ forms a quantized Hall
insulator with Chern number $C_0 = -1$, while each $f_k$ is undergoing a Chern-number changing transition from
Chern numbers $C_k=1$ to $C_k=0$. (\ref{biqhtrans}) is simply the field theory for this transition. 
Since the spinful fermions are undergoing a Chern-number changing transition, the gap to spinful excitations 
closes at this transition. Note that the direct transition involving two Dirac fermions is protected by the $SU(2)$ spin symmetry; when it is broken,
and there is no additional symmetry protecting $M_\uparrow = M_\downarrow$,
the direct transition will split into two transitions. The charge-$1$ superfluid phase of $b_k$ requires 
$(C_\uparrow, C_\downarrow) = (1,0)$ or $(0,1)$, and therefore requires the $SU(2)$ to be broken. 
Note that in this theory, it is important that there is no chemical potential term for the fermions, which 
requires that the density of $f_\alpha$, and correspondingly $b_\alpha$, each be constant through the transition. 

Now consider the above theory, but with the global $U(1)$ symmetry associated with $A_e$ 
explicitly broken to $Z_2$ by a charge-2 Higgs scalar:
\begin{align}
\mathcal{L} = &|(\partial - i 2 A_e) \Phi|^2 - m |\Phi|^2 + \lambda |\Phi|^2 
- \frac{1}{4\pi} A_e \partial A_e
\nonumber \\
&+\sum_{k=\uparrow, \downarrow} \bar{\psi}_k [\gamma^\mu (\partial_\mu - i \alpha_\mu) + M_k] \psi_k
- \frac{1}{2\pi} A_e \partial \alpha ,
\end{align}
with $m > 0$ so that $\Phi$ is condensed: $\langle \Phi \rangle \neq 0$. The insulating states
are now replaced by gapped states with a $Z_2$ global symmetry. Remarkably, the BIQH state descends
into the topologically non-trivial $Z_2$ SPT state, while the trivial insulator descends into
the topologically trivial $Z_2$ symmetric gapped state. The superfluid descends into a
$Z_2$ symmetry breaking state. This can be seen as follows. The BIQH state has two counterpropagating
edge modes, described by two chiral boson fields $\phi_L$ and $\phi_R$. The backscattering term
$\cos(a(\phi_L + \phi_R))$, for integer $a$, is prohibited by the $U(1)$ charge conservation, because
only $\phi_L$ transforms under the action of the $U(1)$ symmetry. If $U(1)$ is broken to $Z_2$, terms
of the form $\cos(2a(\phi_L + \phi_R))$ are now allowed in the edge Hamiltonian as they preserve the $Z_2$ symmetry.
However if they generate an energy gap, then $\langle e^{i (\phi_L + \phi_R) } \rangle \neq 0$, which breaks
the $Z_2$ symmetry. Therefore the edge states still cannot be gapped without breaking the $Z_2$ symmetry.
This is the signature of the topologically non-trivial $Z_2$ SPT state. In contrast, the trivial Mott insulator 
has no edge states in the presence of either the $U(1)$ or $Z_2$ global symmetries. 

The critical theory between the $Z_2$ trivial and non-trivial SPT states becomes a critical
theory between the doubled semion and toric code models when the global $Z_2$ symmetry is gauged. 
This can be done in the above theory by replacing $A_e$ with a dynamical $U(1)$ gauge field $a$:
\begin{align}
\mathcal{L} = &|(\partial - i 2 a) \Phi|^2 - m |\Phi|^2 + \lambda |\Phi|^2 
- \frac{1}{4\pi} a \partial a
\nonumber \\
&+\sum_{k=\uparrow, \downarrow} \bar{\psi}_k [\gamma^\mu (\partial_\mu - i \alpha_\mu) + M_k] \psi_k  - \frac{1}{2\pi} a \partial \alpha ,
\end{align}
This occurs when we interpret $b_k$ as Schwinger bosons (which are related to the $SU(2)$ spins by 
$\vec{S} = \frac12 b^\dagger \vec{\sigma} b$). 

When $\Phi$ is condensed, at long wavelengths the effective theory can be written in terms of the phase $\theta$ of $\Phi$:
\begin{align}
\mathcal{L} = &|(\partial \theta - 2 i a)|^2 + \sum_{k=\uparrow, \downarrow} \bar{\psi}_k [\gamma^\mu (\partial_\mu - i \alpha_\mu) + M_k] \psi_k
\nonumber \\
&- \frac{1}{4\pi} a \partial a - \frac{1}{2\pi} a \partial \alpha 
\end{align}
Picking the gauge $\theta = 0$, $a_0 = 0$ and integrating out $a$ then just gives the action of $QED_3$, with two flavors of Dirac
fermions ($N_f = 2$):
\begin{align}
\mathcal{L} = &\sum_{k=\uparrow, \downarrow} \bar{\psi}_k [\gamma^\mu (\partial_\mu - i \alpha_\mu) + M_k] \psi_k - \frac{1}{g^2} (\epsilon_{\mu\nu\lambda} \partial_\nu \alpha_\lambda)^2
\end{align}
In the large $N_f$ limit, it is known that a critical fixed point exists, although this has not been fully
established when $N_f = 2$. 

\section{Projective construction of BIQH state}
\label{biqhProjConApp}

Here we fill in some details about why the construction presented in the main text
describes the BIQH state. Recall that we set
\begin{align}
b = f_1 f_2,
\end{align}
and we consider a mean-field ansatz where $f_1$, $f_2$ form quantized Hall insulators with Chern
numbers $(C_1, C_2) = (-1,2)$. The emergent $U(1)$ gauge field $a$ is associated with the gauge
redundancy $f_1 \rightarrow e^{i\theta} f_1$ and $f_2 \rightarrow e^{-i\theta} f_2$. Without loss of
generality, suppose $f_1$, $f_2$ carry charges $1$ and $0$, respectively, under the external gauge field
$A_e$. Integrating out $f_1$ and $f_2$ then yields the following effective theory for the gauge fields:
\begin{align}
\mathcal{L} &= -\frac{1}{4\pi} (A_e + a)\partial (A_e + a) + \frac{2}{4\pi} a \partial a
\nonumber \\
&= -\frac{1}{4\pi} A_e \partial A_e - \frac{1}{2\pi} A_e \partial a + \frac{1}{4\pi} a \partial a. 
\end{align}
Since CS term for $a$ has unit coefficient, clearly this describes a gapped state with unique ground state degeneracy
on all closed manifolds. The elementary excitations are particles/holes in the $f_1$ and $f_2$ states, which, after
being dressed by $a$ gauge flux due to the CS term, become bosonic excitations. Furthermore, integrating
out $a$ yields the response theory
\begin{align}
\mathcal{L} = -\frac{2}{4\pi} A_e \partial A_e,
\end{align}
which shows that the state has Hall conductance $\bar{\sigma}_{xy} = 2$. Therefore this state describes a 
BIQH state, with no intrinsic topological order.

An alternative way of seeing this result, and directly deriving the effective theory (\ref{BIQHLag1}), is as follows. 
We introduce three gauge fields, $a^1$, $a^2$, and $a^3$, and their associated conserved currents
$j^I_{\mu} = \frac{1}{2\pi} \epsilon_{\mu\nu\lambda} \partial_\nu a^I_\lambda$, for $I = 1,.., 3$. 
$j^1$ is taken to describe the current of $f_1$. The Chern number $2$ state of 
$f_2$ is then assumed to consist of two filled bands, each with Chern number $1$, so that
the current of $f_2$ can be described by two conserved currents, $j^2$ and $j^3$, each
of which describes the dynamics of a Chern number $1$ band. Now, since $j^I$ each describe
the dynamics of a band with unit Chern number, the effective theory is:
\begin{align}
\mathcal{L} =& \frac{1}{4\pi} (a^2 \partial a^2 + a^3 \partial a^3 - a^1 \partial a^1) + 
\nonumber \\
&\frac{1}{2\pi} a \partial (a^2 + a^3 - a^1) + \frac{1}{2\pi} A_e \partial a^1.
\end{align}
Integrating out $a$ then gives (\ref{BIQHLag1}), for $n = 1$. 

Finally, note that the case where $(C_1, C_2) = (-k,k+1)$ also describes a 
BIQH state, with Hall conductance $\bar{\sigma}_{xy} = k(k+1)$. 

\section{Physical operators at the CSL - $Z_2$ critical point}
\label{physicalOps}

Recall the critical theory between the $1/2n$ Laughlin FQH state and the $Z_2$ fractionalized states
is described by
\begin{align}
\label{CSHiggsv}
\mathcal{L} = &|(\partial - i 2 A) \Phi|^2 + m |\Phi|^2 + \lambda |\Phi|^2 
\nonumber \\
&- \frac{2n}{4\pi} A \partial A + \frac{1}{2\pi} A_e \partial A.
\end{align}
In the case of the FQH transition, where the physical degrees of freedom are bosons with a conserved
$U(1)$ charge, the boson destruction operator at the critical point is
$b = \hat{M} \Phi$,
where $\hat{M}$ removes $2\pi$ units of flux of $A$. Due to the CS term, $2\pi$ flux of $A$ carries 2 units of 
$A$ charge, and therefore the physical gauge-invariant operator must also remove a quanta of $\Phi$. 
From the coupling of the external field, we see that a $2\pi$ flux of $A$ will carry
unit charge under $A_e$ and therefore corresponds to the physical boson. 
$\Phi$ describes double vortices, as it carries charge $2$ under $A$. 
Single vortices remain gapped through the transition and do not appear in the low energy theory at the critical point. 

Now let us consider the $SU(2)$ invariant spin-$1/2$ system. Here, the critical theory is described by 
(\ref{CSHiggsv}), but without the external gauge field $A_e$ and with $A$ replaced by the gauge field $a$ to which the Schwinger bosons couple.
The Higgs field $\Phi$ itself can be physically understood as a spin singlet pair of the Schwinger bosons. 
The single spin operators do not appear as scaling operators in the critical theory. 
The fact that the partons $f_\alpha$ remain in a gapped Chern insulator
indicates that the triplet gap remains finite through the transition. The only 
scaling operators in the theory are therefore spin singlet operators. These include the gauge flux:
$j_\mu \equiv \frac{1}{2\pi}\epsilon_{\mu \nu \lambda} \partial_\nu a_\lambda \propto
\epsilon_{\mu\nu\lambda} \vec{S} \cdot (\partial_\nu \vec{S} \times \partial_\lambda \vec{S})$.
$\int d^2 r j_0$ is proportional to the skyrmion number of $\vec{S}$. 
The other basic gauge invariant operator is $\hat{M} \Phi$, 
which physically corresponds to the operator that changes the skyrmion number by 1. 
The scaling functions of these operators can be readily obtained through various large $N$ approximations. These
were performed for $U(1)$ CS-Higgs theories in \cite{wen1993}, where it was shown that the transitions
are indeed continuous in the large $N$ limit. 

\section{CSL - $Z_2$ transition for general $n$}
\label{generalN}

In the main text we discussed the transition between the $n =1$ CSL and the $Z_2$ spin liquid (doubled semion). 
To describe the case with general $n$, we start with the $n$ flavors of two-component Schwinger bosons:
$b_\alpha(\b{r}) \approx \sum_{\beta=1}^n e^{i\b{Q}_\beta \cdot \b{r}}b_{\alpha \beta}(\b{r})$ 
for $\beta = 1,...,n$ and $\alpha =\uparrow, \downarrow$, and we consider a state where one of the $n$ flavors 
transitions from the $\bar{\sigma}_{xy} =2$ BIQH to the charge-2 SC while the other $n-1$ flavors stay in the $\bar{\sigma}_{xy} = 2$ BIQH state. 
The resulting state is described by a $U(1) \times U(1)$ CS theory with $K = \left(\begin{matrix} 0 & 2 \\ 2 & 2n \end{matrix}\right)$,
which is topologically equivalent to the toric code/doubled semion models when $n$ is even/odd, respectively. 
The wave function that interpolates through this transition, for general $n$, is given by (\ref{wfnPg}), with $|\Phi_{BIQH} \rangle$ replaced by 
$ \sum_\beta e^{-i Q_\beta \cdot r} \langle 0| f^\dagger_{1 \beta} f^\dagger_{\alpha \beta} | \Phi_{fMF} \rangle$,
where $|\Phi_{fMF}\rangle$ is the mean-field state of the fermionic partons $f_{1,\beta}$ and $f_{\alpha \beta}$ 
where $f_{\alpha \beta}$ for $\beta = 1, ...,n$ form spin singlet Chern insulators with Chern number $2$,
and $f_{1\beta}$ form a Chern insulator with Chern number $-1$, while one of the $f_{1 \beta}$ undergoes
a Chern number changing transition from Chern number $-1$ to $-2$. $|0\rangle$ denotes the $f$-vacuum.

\section{Review of fermionization of 3D XY transition}
\label{xyAppendix}

Here we will briefly review the fermionization of the 3D XY transition, which was proposed in \cite{chen1993}.
It will be helpful to consider Table \ref{statesTable} of the main text. Notice that the Chern number assignment
$(C_1, C_2) = (1,0)$ for the mean-field states of $f_1$ and $f_2$ leads to a description of a topologically trivial
Bose Mott insulator. In contrast, the case $(C_1, C_2) = (1,-1)$ describes the Bose superfluid. Therefore, the effective
theory describing the transition between a Bose Mott insulator and a superfluid can be understood as a Chern number
changing transition for $f_2$. Such a critical theory is described by the following action:
\begin{align}
\label{fermL}
\mathcal{L}_{ferm} = &\frac{1}{4\pi} \frac{1}{2} a \partial a + \bar{\psi} \gamma^\mu (\partial_\mu - i a_\mu) + m \bar{\psi} \psi
\nonumber \\
&+ \frac{1}{4\pi} A_e \partial A_e + \frac{1}{2\pi} A_e \partial a,
\end{align}
where $m > 0$ describes the Mott insulator and $m < 0$ describes the superfluid. 
Since this theory descibes the Mott insulator - superfluid transition in the presence of particle-hole symmetry, it is conjectured
to be equivalent to the conventional 3D XY critical point:
\begin{align}
\label{xyL}
\mathcal{L}_{xy} = |(\partial - i A_e) \Phi|^2 + m |\Phi|^2 + \lambda |\Phi|^4
\end{align} 
Rescaling $A_e \rightarrow A_e' = A_e/2$ and subtracting both (\ref{fermL}) and (\ref{xyL}) by $\frac{2}{4\pi} A_e \partial A_e$ gives the duality used in the
main text.

\section{Insufficiency of slave fermion construction}
\label{slaveFermApp}

In this section, we discuss in some more detail the necessity of the BIQH construction of this paper, as compared
with the fermionic IQH construction of \cite{wen1989}, for understanding the transitions between the CSL and the 
$Z_2$ spin liquids. The construction of \cite{wen1989} starts by writing the spin-1/2 operator in terms of slave
fermions:
\begin{align}
\vec{S} = \frac12 c^\dagger \vec{\sigma} c,
\end{align}
where $c^\dagger = (c_\uparrow^\dagger, c_\downarrow^\dagger)$ is a two-component fermion. This leads to an emergent
$U(1)$ gauge field $a$ associated with the gauge transformation $c \rightarrow e^{i\theta} c$. The CSL corresponds
to a state where the fermions $c$ form a spin singlet Chern insulator with Chern number $C_c = -2n$.
For $n = 1$, this leads to a state that is equivalent to the KL-CSL, while for $n > 1$, the fractional statistics of the 
quasiparticles are slightly different from the KL-CSL, as summarized in the main text. 

In order to describe a transition to a $Z_2$ state, one possibility is to consider a transition where the
pair field of the fermions, $\Phi = f_\uparrow f_\downarrow$, condenses, thus breaking the $U(1)$ gauge symmetry to
$Z_2$. In contrast to the BIQH state, in the fermionic case there is no theory of a transition from a fermion IQH state 
that simultaneously condenses the pair field and completely destroys the edge modes. Instead we can consider a scenario where the pair field condenses,
but the edge modes are not destroyed. To understand the nature of the resulting state, let us focus on the case
$n  =1$. The IQH states of the fermions can be described by two $U(1)$ CS gauge fields $a^\uparrow$, $a^\downarrow$, 
such that $j_{\alpha;\mu} = \frac{1}{2\pi} \epsilon_{\mu\nu\lambda} \partial_\nu a_\lambda^\alpha$ describes the current of
$c_\alpha$. The effective theory is therefore:
\begin{align}
\mathcal{L} = &\frac{1}{4\pi} (a^\uparrow \partial a^\uparrow + a^\downarrow \partial a^\downarrow) + \frac{1}{2\pi} a \partial (a^\uparrow + a^\downarrow) + 
\nonumber \\
&|(\partial - 2 i a) \Phi|^2 + m |\Phi|^2 + \lambda |\Phi|^4. 
\end{align}
Here, $\Phi$ represents the pair field $f_\uparrow f_\downarrow$; its condensation breaks the $U(1)$ gauge symmetry 
to $Z_2$. In order to understand the topological properties of the $\Phi$-condensed phase, it is helpful to 
perform a particle-vortex duality on $\Phi$:
\begin{align}
\mathcal{L} = &\frac{1}{4\pi} (a^\uparrow \partial a^\uparrow + a^\downarrow \partial a^\downarrow) + 
\frac{1}{2\pi} a \partial (a^\uparrow + a^\downarrow + 2\alpha) + 
\nonumber \\
&|(\partial - i \alpha) \Phi_v|^2 + m' |\Phi_v|^2 + \lambda' |\Phi_v|^4. 
\end{align}
Here, $j_{\Phi;\mu} = \frac{1}{2\pi} \epsilon_{\mu\nu\lambda} \partial_\nu \alpha_\lambda$ is the current of the original
$\Phi$, while $\Phi_v$ describes the vortices of $\Phi$. The phase where $\Phi$ is condensed corresponds to the
case where $\Phi_v$ is uncondensed, and vice versa. Therefore, when $\Phi$ is condensed, $\Phi_v$ is uncondensed
and can be integrated out, leaving us with a $U(1)^4$ CS theory with $K$-matrix
\begin{align}
K = \left(\begin{matrix} 
1 & 0 & 1 & 0 \\
0 & 1 & 1 & 0 \\
1 & 1 & 0 & 2 \\
0 & 0 & 2 & 0 \\
\end{matrix}\right).
\end{align}
This satisfies $|\text{Det } K| = 4$. However it has three positive eigenvalues and one negative eigenvalue, implying that the
system has a net chiral central charge $c = 2$, and therefore has topologically protected gapless edge states. Such a state
is topologically distinct from both the toric code and doubled semion models. It can be thought of as a chiral spin liquid
with a discrete gauge structure, \it eg \rm a chiral topological superconductor with $4$ chiral edge Majorana modes coupled to a
fluctuating $Z_2$ gauge field. Such a state can be constructed using the honeycomb Kitaev model \cite{kitaev2006,yao2007}. 

While it is interesting that this ``$Z_2$ CSL'' also neighbors the KL-CSL, it does not correspond to the two different $Z_2$ spin liquids
considered in the main text. It appears that the construction in terms of fermionic spinons \cite{wen1989} is indeed insufficient
to describe the transition between the CSL and the $Z_2$ spin liquids, although it does allow access to the transition to
a $Z_2$ CSL. 

\section{Topological properties of Abelian Chern-Simons-Higgs theories}
\label{CSHapp}

Here we will briefly review the topological properties of Abelian CS-Higgs theories \cite{dijkgraaf1990}(see also \cite{cheng2013} for a recent
discussion). Specifically, consider a $U(1)_{n}$ CS term, coupled to a charge-$k$ Higgs field:
\begin{align}
\label{CSH}
\mathcal{L} = |(\partial - i k A)\Phi|^2 + m|\Phi|^2 + \lambda |\Phi|^2 + \frac{n}{4\pi} A \partial A,
\end{align}
where $\lambda > 0$. When $\Phi$ is uncondensed this describes $U(1)_n$ CS theory, the topological
properties of which are well known \cite{wen04}: there are $n$ topologically distinct quasiparticles, with fractional statistics
$\theta_a = \pi a^2/n$. 

In order to understand the topological properties of the Higgs phase, where $\Phi$ is condensed and the
gauge group $U(1)$ is broken to $Z_k$, it is helpful to perform a duality transformation on $\Phi$ and 
to consider the theory in terms of the vortices of $\Phi$. This is described by the theory
\begin{align}
\mathcal{L} = &|(\partial - i \alpha)\Phi_v|^2 + \bar{m}|\Phi_v|^2 + \bar{\lambda} |\Phi_v|^2 
\nonumber \\
&+ \frac{k}{2\pi} A \partial \alpha + \frac{n}{4\pi} A \partial A,
\end{align}
where $\Phi_v$ is a complex scalar describing the vortices of $\Phi$. The condensed phase of $\Phi$ is therefore
the uncondensed phase of $\Phi_v$. Considering the case where $\Phi_v$ is uncondensed (Higgs phase of (\ref{CSH})), we can integrate it 
out to obtain a $U(1) \times U(1)$ CS theory:
\begin{align}
\mathcal{L} = \frac{n}{4\pi} A \partial A + \frac{k}{2\pi} A \partial \alpha. 
\end{align}
The topological properties of such a theory can be directly read off from the $K$-matrix:
\begin{align}
K = \left(\begin{matrix}
n & k \\
k & 0 \\
\end{matrix} \right).
\end{align} 
It has $k^2$ distinct quasiparticles. For $n$ even, this describes a topological phase 
where the local degrees of freedom are all bosons, and otherwise the
local degrees of freedom contain fermions. In general, we can
perform an $SL(2;Z)$ transformation $K \rightarrow K' = W^T K W$, with $W \in SL(2;Z)$, 
which keeps the topological properties of $K$ invariant, such that 
\begin{align}
K' = \left(\begin{matrix}
n -2k & k \\
k & 0
\end{matrix} \right).
\end{align}
Therefore, the case where $n$ is a multiple of $2k$ leads to 
$K' = \left(\begin{matrix}
0 & k \\
k & 0
\end{matrix} \right)$,
which encodes the topological properties of ordinary $Z_k$ gauge theory. 

When $n$ is even, the above can be viewed as a generalized version of $Z_k$ gauge theory that
corresponds to one of the $k$ different kinds of $Z_k$ gauge theory found in 
\cite{dijkgraaf1990}.

The case $n = k = 2$ describes the doubled-semion model discussed in the main text.

\end{document}